\documentclass[twoside]{article}

\usepackage{PRIMEarxiv}

\usepackage[utf8]{inputenc} % allow utf-8 input
\usepackage[T1]{fontenc}    % use 8-bit T1 fonts
\usepackage{hyperref}       % hyperlinks
\usepackage{url}            % simple URL typesetting
\usepackage{booktabs}       % professional-quality tables
\usepackage{amsfonts}       % blackboard math symbols
\usepackage{nicefrac}       % compact symbols for 1/2, etc.
\usepackage{microtype}      % microtypography
\usepackage{lipsum}
\usepackage{fancyhdr}       % header
\usepackage{tabularx}
\usepackage{multirow}
\usepackage{graphicx}       % graphics
\usepackage{natbib}
\graphicspath{{media/}}     % organize your images and other figures under media/ folder

\title{Precarity and solidarity: Preliminary results on a study of queer and disabled fiction writers' experiences with generative AI
}

\author{
    C.E. Lamb \\
    School of Computing \\
    Queen's University \\
    Kingston, Ontario, Canada \\
    \texttt{cel4@queensu.ca} \\
    \And
    D.G. Brown, M.R. Grossman \\
    Cheriton School of Computer Science \\
    University of Waterloo \\
    Waterloo, Ontario, Canada \\
    \texttt{\{dan.brown,maura.grossman\}@uwaterloo.ca} \\
}

\begin{document}
\maketitle

\begin{abstract}
We have undertaken a mixed-methods study of fiction writers' experiences and attitudes with generative AI, primarily focused on the experiences of queer and disabled writers. We find that queer and disabled writers are markedly more pessimistic than non-queer and non-disabled writers about the impact of AI on their industry, although pessimism is the majority attitude for both groups. We explore ways that generative AI exacerbates existing sources of instability and precarity in the publishing industry, reasons why writers are philosophically opposed to its use, and individual and collective strategies used by marginalized fiction writers to safeguard their industry from harms associated with generative AI.
\end{abstract}

% keywords can be removed

\section{Introduction}

Since the recent popularization of large language models (LLMs) and text-to-image models, a great deal of ink has been spilled on predicting their effects on various creative industries - either revolutionizing and democratizing these industries or exploiting human artists and spelling the death of creativity. In order to bring more evidence-based arguments into this space, we performed a mixed-methods survey of 137 professional prose fiction writers asking about their experiences with generative AI (genAI) and how it has affected their careers. The survey covers not only writers' direct experiences with using AI but also broader questions about the recent past, present, and future of the industry. We used descriptive statistics for the quantitative questions in the survey and a grounded theory approach for analyzing the writers' qualitative, free-text responses, providing outside sources to further contextualize writers' responses where necessary.

Our primary research question is: What effect is genAI currently having on the careers of professional fiction writers? And our secondary question, following naturally from the primary question, is: What strategies do professional fiction writers employ to navigate and mange these effects?

The sampling method for our study resulted in a reasonably representative sample of writers in some areas, but significantly overrepresented writers who identify as queer and/or disabled. Rather than try to correct this overrepresentation, we found it intriguing and have decided to begin our analysis by centering queer and disabled writers' experiences specifically.

Quantitatively, our data shows that queer and disabled writers are more precarious and more vulnerable to the potential negative effects of genAI than other writers - with significantly less median writing income despite similar years of experience in the industry. As a result, they are more uniformly pessimistic about the effects of genAI than non-queer and non-disabled writers, although pessimism pervades the majority of both groups.

Qualitatively, the queer and disabled writers whose text responses we have analyzed provide a coherent account of their experiences and attitudes regarding genAI, with precarity being the central concept. Writers' attitudes towards genAI must be understood in context of the writers' career goals and reasons for writing and of the significant precarity that already exists in their industry. The presence of genAI worsens most aspects of this existing precarity. Moreover, the writers surveyed view genAI both as inherently unhelpful for their goals and as ethically unacceptable because of the unauthorized use of copyrighted training data. Because genAI is viewed as unhelpful, unethical, and actively dangerous to the writing industry, marginalized writers engage in collective strategies to discourage the use of genAI in publishing through all informal means available to them.

In addition to providing first-hand qualitative accounts of a relevant population's experience with AI, our study demonstrates the importance of understanding experiences with AI, not merely as a series of interactions with software, but in terms of their broader social context. In the context of fiction writing, backlash against genAI should be viewed not as the reflexive result of polarization or resistance to change, but as a conscious and self-protective strategy engaged in by a vulnerable population of artists.

In light of these results, we conclude by offering a few brief policy suggestions for how human writers' copyrights, and their best interests more broadly, can be protected in the face of technological change.

\section{Related Work}

Our study is not the first scholarly study of the effect of genAI on human artists. Even before the popularization in 2022 of LLMs and text-to-image generators (referred to, along with similar models in music and other arts, as ``genAI''), the field of computational creativity often produced and studied co-creative experiences with both professional creative humans and computers in the loop (e.g., \citet{colton2016beyond}). Since 2022, qualitative studies of professional artists working with genAI include \citet{kaila2024gardening}, who studied the frictions encountered by artists in varied fields when using genAI; \citet{vimpari2023adapt}, who used template analysis on interviews with game industry professionals about their use of text-to-image generation AI; and \citet{uusitalo2024clay}, who used template analysis to study the use of genAI by UX and industrial designers. However, although both fields are creative, the workflows and professional needs of prose fiction writers - working individually and freelance, for example, rather than in teams - are in many ways different from the workflows and needs of UX designers.

\citet{ippolito2022creative}, at Google Research, performed an in-depth user study of 13 professional writers and their interactions with a prototype story generation system. This study gives a great deal of useful information about the actual experience of using LLMs for creative writing, the interaction patterns that emerge, and the advantages and difficulties that these tools can introduce into a professional writer's process. However, the writers in the study are volunteers who agreed to participate in the study as an experiment, and they did so prior to the widespread popularization of LLMs; the broader impact of LLMs on the writing industry is beyond the study's scope.

\citet{jiang2023ai} is also relevant to our work; it is a review and position paper summarizing harms to professional visual artists from the proliferation of genAI. It includes professional visual artists and prominent AI ethicists among its authors. Several of \citet{jiang2023ai}'s themes agree with those brought up by the writers in our study, including significant concern about plagiarism and copyright infringement as well as a philosophical argument - closely resembling our writers' in situ concept of ``soul,'' described below - against viewing genAI output as ``art.''

We would be remiss not to mention several non-scholarly surveys that have gathered fiction writers' opinions about genAI. The entities producing these surveys include established groups representing writers' interests, such as International Thriller Writers (\cite{international2023artificial}) and The Author's Guild (\cite{authors2023survey}) - as well as the group Humanity In Fiction, which was created specifically in response to the rise of LLMs (\cite{humanity2023authors}). These surveys each revealed significant concern among professional fiction writers, both about ethical issues such as the use of copyrighted data and about the overall effects of genAI on their industry. With our study we intend to add a qualitative element to these results, emphasizing \textit{why} fiction writers have the opinions they do and what they are doing about it, as well as how marginalized writers appear to be disproportionately affected; we also aim to make these ideas more visible to computer scientists studying or actively working on the design of AI.

\begin{table}
  \caption{Demographics of our data - both the whole dataset, and the two sets into which it was split. The Author's Guild survey (\cite{authors2023key}) is used as a comparison, for ``ground truth'' about the overall demographics of professional fiction writers. Numbers do not always add up to 100\% as not all writers answered all demographic questions, and some response options are omitted for clarity.}
   \centering
   \begin{tabularx}{\textwidth}{XXXXX}
     \toprule
     & Whole Dataset & Non-Queer and/or Non-Disabled & Queer and Disabled & Author's Guild \\
     Average Age & 42 & 44.5 & 38.6 & - \\
     Gender & 49\% female \newline 22\% male \newline 23\% nonbinary & 57\% female \newline 26\% male \newline 12\% nonbinary & 44\% female \newline 23\% male \newline 26\% nonbinary & 6\% nonbinary \\
     Race & 85\% white & 87\% white & 79\% white & 79\% white \\
     Disabled & 62\% & 18\% & 100\% & 11\% \\
     LGBTQ+ & 63\% & 26\% & 100\% & 12\% \\
     Employment & 23\% write full time & 26\% write full time & 13\% write full time & 35\% write full time \\
     Average years of writing experience & 11.2 & 11.75 & 10.3 & - \\
     Median writing income & \$1500 & \$3000 & \$695.50 & \$2000 \\
     Mean writing income & \$14,665.16 & \$18,657.65 & \$7012.33 & - \\
     \bottomrule
   \end{tabularx}
   \label{tab:demographic}
 \end{table}

\section{Methods and Data Collection}

This research study was approved by the Queen's University General Research Ethics Board. After receiving approval, we conducted our survey in May 2024. We recruited professional fiction writers using snowball sampling, beginning by advertising in writers' Slacks and Discords and on the Bluesky account of our first author, who is a professional fiction writer themselves. We defined a ``professional fiction writer'' as anyone, anywhere in the world, who has ever received any amount of money for a work of prose fiction. We avoided a more restrictive definition because, unlike in other ``professional'' fields, the majority of published fiction writers do not make enough money from their writing to support themselves (\cite{authors2023key}). We asked questions about each writer's income and years of experience so that we could gauge whether the success and experience of our surveyed writers was comparable to the general population of fiction writers.

Each recruited writer filled out an online questionnaire containing both quantitative and qualitative questions. The quantitative questions asked for demographic information, information on a writer's career, and yes/no and Likert-scale questions about AI, their careers, and the writing industry. The qualitative questions were free-text fields in which writers were invited to elaborate on the answers to the quantitative questions. Writers were able to give or withhold consent for their free-text responses to be quoted directly; even when writers gave consent, we have not included any text responses in this paper which contain identifying information such as the list of books a writer has published. Each writer was given CAD \$20 for participating, in their choice of either a PayPal transfer or an Amazon gift card.

Once the recruitment materials were posted, responses came in much faster than anticipated, and due to an error in implementing the survey software, we received a somewhat larger number of survey responses than anticipated (137 in total). Most of these responses included at least some qualitative as well as quantitative data.

\subsection{Demographics and Representativeness of the Data}

An initial breakdown of our writers by demographic can be seen in Table \ref{tab:demographic}. To assess the representativeness of our data, we compared it to the posted results of the 2023 Author's Guild survey on writer income (\cite{authors2023key}).

This comparison shows that in most respects, our sample of writers is consistent with the general population of writers. Both populations are predominantly white, both are mostly \textit{not} full-time writers (but with a sizeable minority writing full-time), and the average yearly writing income of both sets of writers is of the same order of magnitude. We received responses from writers of various genres such as science fiction, romance, and literary fiction and from writers using various business models (e.g., traditionally published, self-published, or both.) While we did not ask specifically about writers' nationalities, our writers reported income in a variety of currencies, including USD, CAD, GBP, Euros, Australian dollars, and Japanese yen, all of which we converted to USD before calculating the median or mean. However, there are two glaring exceptions. Unlike the general population of fiction writers, 63\% of our writers identify as LGBTQIA+ and 62\% identify as ``neurodivergent or disabled.''\footnote{It may be misleading to compare our number of disabled writers to those in the Author's Guild survey, as the questions are differently phrased in both surveys. The Author's Guild asked if writers ``fit the Americans With Disabilities Act's definition of impairment,'' while we merely asked if writers self-identified as disabled.} We believe these two statistics to be an artifact of our snowball sampling method, as our first author openly identifies as queer and disabled themselves.

Rather than discarding our data as biased, we realized that queer and disabled fiction writers are interesting populations to study, being intersectionally (\cite{crenshaw2013demarginalizing}) marginalized groups within the publishing industry whose voices might be overlooked in existing research. We decided that, in our qualitative analysis, we will first analyze the writers in our data who are both queer and disabled. The qualitative responses of the remaining writers (writers who are non-queer, non-disabled, or both non-queer and non-disabled) will be held in reserve for a later point, where we can investigate whether and in what ways our results on queer and disabled writers generalize to other writers as well. This results in two separate survey subpopulations: a group of 54 queer disabled writers, and a control group of 75 non-queer and/or non-disabled writers. (The remaining 8 writers did not answer enough questions to be categorized into either group and were therefore discarded.)

With regards to income, the median is used. As will be immediately apparent from Table \ref{tab:demographic}, the median income of fiction writers is quite low. This median is taken from a highly skewed distribution, with a mean much higher than the median. The majority of writers make very little money; a few make considerably more than the median, sometimes enough to live on, with a handful (in our data) reaching the low six figures. As should be clear from the comparison with Author's Guild figures, this is not an artifact of our data collection method but is true of the population of fiction writers generally.\footnote{The Author's Guild survey reports several different ``median writing income'' figures. The figure of \$2000 reflects median ``book-related'' income for all authors who took the survey. The figure rises to \$5000 when considering all forms of ``author-related'' income, including ``editing, blogging, teaching, speaking, book coaching, copy writing and journalism,'' and rises still further when only considering full-time writers, or only writers in ``commercial markets.'' Because our survey is not limited to full-time writers or to writers in a particular market or genre, and because we only asked about income earned from fiction writing, the figure of \$2000 is the most appropriate figure for comparison.}

\subsection{Qualitative Analysis}

We completed our qualitative analysis in several steps, using a grounded theory methodology based on \citet{corbin2015basics}. First, all three authors briefly reviewed all the gathered qualitative responses together, including those of non-queer and non-disabled writers. Second, we selected writers to analyze through open coding. While we could not perform formal theoretical sampling due to the mixed-methods nature of the study, we could approximate theoretical sampling by selecting writers based on questions that appeared in our developing theory. The first writer, Writer A, was selected at random and subsequent writers were selected based on specific questions that arose as we discussed the developing concepts. For example, did writers' experiences look different based on whether they were experienced or inexperienced, whether they were marginalized racially or in other ways? When writers' quantitative responses were different from others - for instance, if they showed more optimism than others, or if they indicated the had used genAI when most writers had not - then how were these differences explained in their qualitative responses?

Identifying data such as emails were separated from writers' responses before analysis and the writers are referred to with alphanumeric designations. Five non-queer and/or non-disabled writers - B, F, G, H, and I - were analyzed before we made the decision to focus initially only on queer and disabled writers. These analyses were not discarded but were set aside until we get to the stage of comparing queer and disabled writers' responses to others and they are not included in the current version of our grounded theory.

We repeated this process for a total of 16 queer and disabled writers - A through U, minus the aforementioned five - until we no longer had specific questions that could be easily answered by selecting a particular writer to analyze. At that point, the central shape and concepts of our developing theory were apparent to us. We did not have saturation - each additional writer analyzed still added new insights - but these insights were becoming increasingly minor, adding small new details or connections to existing concepts. Consequently, and considering the time-sensitive, rapidly evolving nature of this research, we wanted to pause here to write up our preliminary results.

After publishing this preprint, we plan to resume our analysis until the responses of every queer and disabled writer in our data have been carefully analyzed and coded. Then, we will analyze the writers in our control group to see whether and in what ways our theory generalizes to non-queer and non-disabled fiction writers.

 \begin{table}
  \caption{Optimism and pessimism about the fiction industry from the core group of queer and disabled writers. Numbers do not sum to 100\%, because not all writers answered all questions.}
   \centering
   \begin{tabularx}{\textwidth}{XXXXXX}
     \toprule
    & \multicolumn{5}{c}{Queer and Disabled} \\
     & Much harder & Somewhat harder & Neither easier nor harder & Somewhat easier & Much easier \\
     Making money (self, past five years) & 8\% & 23\% & 30\% & 15\% & 4\% \\
     Making money (all writers, past five years)  & 23\% & 28\% & 11\% & 0 & 0 \\
     Making money (self, next five years) & 8\% & 34\% & 16\% & 17\% & 0 \\
     Making money (all writers, next five years) & 34\% & 41\% & 13\% & 0 & 0 \\
     Getting publishers' attention (self, past five years) & 6\% & 38\% & 28\% & 21\% & 9\% \\
     Getting publishers' attention (all writers, past five years)  & 23\% & 34\% & 15\% & 0 & 0 \\
     Getting publishers' attention (self, next five years)  & 11\% & 25\% & 23\% & 19\% & 4\% \\
     Getting publishers' attention (all writers, next five years)  & 30\% & 40\% & 13\% & 0 & 0 \\
     Getting readers' attention (self, past five years) & 11\% & 14\% & 23\% & 21\% & 6\%  \\
     Getting readers' attention (all writers, past five years)  & 23\% & 34\% & 21\% & 6\% & 0 \\
     Getting readers' attention (self, next five years)  & 15\% & 25\% & 23\% & 19\% & 2\% \\
     Getting readers' attention (all writers, next five years)  & 28\% & 36\% & 17\% & 2\% & 0 \\
     & Entirely & Mostly & Somewhat & Slightly & Not at all \\
     Changes caused by genAI?  & 0 & 13\% & 58\% & 13\% & 0  \\
     & Very negative & Somewhat negative & Neutral & Somewhat positive & Very positive \\
     Effects of genAI  & 72\% & 8\% & 2\% & 2\% & 0 \\
     & Very significant & Somewhat significant & Neither significant nor insignificant & Somewhat insignificant & Very insignificant \\
    & 19\% & 38\% & 9\% & 53\% & 4\% \\
     \bottomrule
   \end{tabularx}
   \label{tab:queerquant}
 \end{table}

 \begin{table}
  \caption{Optimism and pessimism about the fiction industry from the control group of non-queer and/or non-disabled writers. Numbers do not sum to 100\%, because not all writers answered all questions. P-values represent the result of a Mann-Whitney U-test measuring the probability that these results and those of the queer disabled group come from the same distribution; p-values less than 0.05 are bolded.}
   \centering
   \begin{tabularx}{\textwidth}{XXXXXXX}
     \toprule
    & \multicolumn{5}{c}{Non-Queer and/or Non-Disabled} \\
     & Much harder & Somewhat harder & Neither easier nor harder & Somewhat easier & Much easier & p \\
     Making money (self, past five) & 13\% & 25\% & 20\% & 21\% & 5\% & 0.85 \\
     Making money (all, past five) & 25\% & 32\% & 20\% & 4\% & 3\% & \textbf{0.049} \\
     Making money (self, next five) & 11\% & 36\% & 20\% & 16\% & 7\% & 0.81 \\
     Making money (all, next five) & 27\% & 37\% & 15\% & 4\% & 4\% & \textbf{0.045} \\
     Getting publishers' attention (self, past five) & 13\% & 27\% & 21\% & 15\% & 9\% & 0.20 \\
     Getting publishers' attention (all, past five) & 25\% & 32\% & 23\% & 3\% & 4\% & \textbf{0.033} \\
     Getting publishers' attention (self, next five) & 9\% & 27\% & 24\% & 16\% & 7\% & 0.83 \\
     Getting publishers' attention (all, next five) & 23\% & 36\% & 17\% & 3\% & 4\% & 0.25 \\
     Getting readers' attention (self, past five) & 16\% & 23\% & 12\% & 21\% & 12\% & 0.59 \\
     Getting readers' attention (all, past five) & 16\% & 36\% & 17\% & 12\% & 3\% & 0.2 \\
     Getting readers' attention (self, next five) & 15\% & 16\% & 16\% & 27\% & 11\% & 0.45 \\
     Getting readers' attention (all, next five) & 27\% & 33\% & 15\% & 5\% & 4\% & 0.40 \\
     & Entirely & Mostly & Somewhat & Slightly & Not at all  \\
     Changes caused by genAI? & 1\% & 27\% & 35\% & 13\% & 8\% & 0.50  \\
     & Very negative & Somewhat negative & Neutral & Somewhat positive & Very positive \\
     Effects of genAI & 52\% & 3\% & 19\% & 7\% & 5\% & \textbf{0.0035} \\
     & Very significant & Somewhat significant & Neither significant nor insignificant & Somewhat insignificant & Very insignificant \\
     & 31\% & 33\% & 4\% & 9\% & 8\% & 0.18 \\
     \bottomrule
   \end{tabularx}
   \label{tab:straightquant}
 \end{table}

\section{Quantitative Results}

Quantitative analysis has been performed for both groups and the results can be seen in Tables \ref{tab:demographic}, \ref{tab:queerquant}, and \ref{tab:straightquant}. Table \ref{tab:demographic} shows the demographic breakdown between the queer and disabled group and the remaining writers, while Tables \ref{tab:queerquant} and \ref{tab:straightquant} show writers' responses to the Likert-scale questions of our survey. We verified the Likert-scale results with a Mann-Whitney U-test to evaluate whether the responses of the queer and disabled writers, and the responses of the remaining writers, belonged to the same distribution.

A few things are notable from Table \ref{tab:demographic}. First, despite being of similar ages and having similar years of professional experience, our queer and disabled writers make much less income from writing than our control group and are less likely to be able to write full-time. (This is in partial contrast to the Author's Guild survey, which reported that full-time LGBTQ+ writers make \textit{more} income than average, although full-time disabled writers make less (\cite{authors2023key}).\footnote{The Author's Guild survey analyzes several forms of marginalization in the same question and at the same time, which means, confusingly, that LGBTQ+ writers and disabled writers are contrasted not against straight writers and able-bodied writers but against "white writers."} However, the Author's Guild survey did not discuss the intersection of these two groups, and it did not report how marginalization affects income for non-full-time writers, which our data includes.)

Certain results from Tables \ref{tab:queerquant} and \ref{tab:straightquant} are also visible at a glance. Both groups have a general tendency to pessimism concerning the state of the writing industry and the effects of genAI. But the tendency to pessimism is stronger and more unanimous among queer and disabled writers. These writers are \textit{both} subject to worse-than-average career outcomes, \textit{and} also perceive and expect worse outcomes for their peers. Mann-Whitney U-tests showed that this difference between queer and disabled writers and our control group was significant for four questions: ease of making money, for all writers, in the past five years; ease of making money, for all writers, in the next five years; ease of getting a publisher's attention, for all writers, in the past five years; and whether the anticipated changes caused by genAI in the next five years were negative or positive. On all four of these questions, queer and disabled writers are significantly more pessimistic than the control group.

In particular, only two queer disabled writers - M and N - expressed any optimism at all on any of the questions about writers in general. Even in these two cases, the optimism was mixed (appearing on certain questions only.) Although M and N are not as opposed to genAI as their peers, their qualitative answers indicate that they broadly agree on the most major points described below - that writers need to be compensated when their work is used as training data, that LLMs are useful to them only in limited ways, and that there are aspects of a human's conscious experience in writing fiction that current AI cannot replicate.

Curiously, despite dramatically lower self-reported income, queer and disabled writers are as likely as the control group to report that their own individual career has become easier in the past five years, or that it will become easier in the next five. Their pessimism instead appears most strongly when considering the state of the industry as a whole.

\section{Preliminary Qualitative Results}

This theory is tentative and still needs to be developed further by analyzing our remaining writers' responses until saturation. However, our queer and disabled writers are strongly in agreement with each other on many points, and these points have a clear conceptual relationship to each other; consequently, our confidence in this preliminary theory is strong.

The central concept of our grounded theory is ``precarity.'' Fiction writing is a deeply precarious industry in ways in which people outside the industry, including computer scientists and software developers, are often unaware of. A great deal of what our writers wanted to tell us had to do with precarious conditions in the industry that had existed before the rise of genAI. This information was not irrelevant - it served to put the writers' opinions about genAI into context. It is necessary to understand this context before explaining the ways genAI has worsened writers' experiences of precarity and their fears for the future.

Knowing the context of the writing industry also helps to explain writers' motivations for writing, which are not primarily financial but have to do with intentional, personal expression and communication. Even though their output is fictional, writers see themselves as expressing something that is unique to them. These motivations lead to two primary objections to genAI. First, because such AI does not share a writer's personal experiences, writers do not see it as useful for helping them to express themselves. Second, because a writer's already-precarious income is tied to ownership of copyright over their work, writers are highly sensitive to the issue of copyrighted work being used as training data without permission; this use is overwhelmingly viewed as plagiarism. Therefore, to marginalized writers in a precarious industry, genAI is viewed as neither useful nor ethical.

Because genAI worsens their already-precarious working conditions, and because it is seen as neither useful nor ethical, queer and disabled writers oppose its use. Rather than passively accepting technological change, these writers employ techniques of boundary-setting and social advocacy in an active attempt to exclude genAI from their industry.

In the following sections, our more detailed results are divided into three sections: existing context prior to genAI, effect of genAI on writers, and strategies employed by writers in response to genAI.

\section{Existing Context}

\subsection{For Love and Money: Motivations For Writing}

Writers in our study report that they write because they love writing. This love consists of three components: an emotional attachment to the activity of writing, a sense of purpose in communicating something personally important, and a desire to connect meaningfully with a reader. In all three of these respects, simply causing a piece of writing to exist is not sufficient. Instead, there is careful intentionality in deciding what to communicate and in determining the most effective way to do so. The enjoyment that a writer gets out of writing comes from this process of intentional, often effortful decision making.

While writers may write out of love, they are also aware that writing is time-consuming and effortful. In order to consistently engage in this activity, they need financial stability. Therefore, writers pay great attention to the financial state of their careers and to the milestones they have reached, or plan to reach, each of which brings them one small step closer to the stability they desire.

\begin{table}
  \caption{Changes in the fiction-publishing industry before and after genAI. The provenance of these changes varies; some are empirically verifiable, while some are the perception and interpretation of the writers in our data.}
   \centering
   \begin{tabularx}{\textwidth}{lXX}
     \toprule
     & Before & After \\
     \hline
     Low Pay & Book advances and short-story payments are generally low and are not keeping up with inflation. Most writers cannot financially support themselves. & Worry over further monetary devaluation of writers' work, due to the perception that ``a machine can do it.'' \\
     \hline
     Flooding the Market & In self-publishing and short fiction, increased number of writers and books competing for the same pool of readers, due in part to new writers beginning careers during COVID-19. & Flooding dramatically increases. Self-publishing and short-fiction markets fill with large amounts of low-quality AI writing. \\
     \hline
     Publishing Opportunities & Shrinking number of available publishers, editors, and marketing capacity per project. & More publishers close or slow down due to the influx of genAI submissions. Writers become unwilling to submit to some publishers due to their use of AI, shrinking the number even further. \\
     \hline
     Digital Marketing & Writers are expected to do their marketing themselves, primarily through social media; useful social media sites for book marketing change to become useless (e.g., Twitter) or adopt pay-for-visibility models. & Trend continues. Some social media platforms also become undesirable due to use of AI or scraping of users' posts for AI training. \\
     \hline
     Preference For Familiarity & In self-publishing, readers prefer writers with known brands. & Preference for known writers increases, especially in self-publishing, due to a suspicion that any unknown writer might be AI. Harder for new writers to establish themselves. \\
     \hline
     Desirability & In traditional publishing, publishers look for ``sure things,'' formulas, and known successes rather than taking artistic risks. Writers feel undesirable to publishers for reasons unrelated to the quality of their book and largely beyond their control. & Publishers attempt to use AI to evaluate submissions. Writers are increasingly evaluated on criteria that they do not control or understand. \\
     \hline
     COVID-19 & Supply chains, networking opportunities, and other aspects of business are disrupted by COVID-19. & Trend continues \\
     \bottomrule
   \end{tabularx}
   \label{tab:before-after}
 \end{table}

\subsection{Existing Instability}

Although writers desire stability, it is extremely challenging to achieve. Instead, the fiction-writing industry is significantly and increasingly precarious. The precise form of this precarity depends on which sector of the industry an writer writes in. The first column of Table \ref{tab:before-after} shows several sources of precarity described by our writers, especially those that have been exacerbated by the use of genAI.

It may help to define, for readers who are unfamiliar, some terms for existing practices in the publishing industry. One form of publishing is known as ``traditional publishing.'' In traditional publishing, writers write book manuscripts which they then send to literary agents. If a writer is selected by an agent, the agent then forwards their manuscripts to editors working at publishing companies. If a publisher purchases rights to the manuscript, then the publisher assumes responsibility for turning the manuscript to a finished book that can be sold by booksellers. This process includes providing editorial guidance where necessary, commissioning cover art and designing a cover, formatting the book for a printer, dealing with a distributor to move the printed books into bookstores, and creating advertising and marketing materials for potential readers. Importantly, copyright of the book text remains with the writer. Rather than purchasing copyright, a publisher purchases rights to publish the book, which are often limited to a specific geographic region and/or a period of time. In exchange for these rights, the writer receives a certain amount of money as an advance on royalties, usually split into several installments. If more than a certain number of copies of the book are purchased by readers, then the writer ``earns out'' the advance and begins to receive additional royalties; otherwise, the advance is the only payment. Rights can revert to the writer if certain conditions are met, such as the book no longer being in print. Sometimes the rights sold to a publisher also include subsidiary rights such as the right to create an audiobook or a translation into another language; these rights can be exercised by the publisher or sold to other companies that specialize in audiobooks, translations, etc.

An alternative model is self-publishing. In self-publishing, a writer assumes all the responsibilities that would normally fall to a publisher - including finding editors, commissioning a cover, formatting, and marketing. The book is then submitted to one or more online retailers, where it can be purchased by readers as an ebook or, in some cases, printed on demand. Overwhelmingly, at least one of the retailers involved is Amazon, which sold over 90\% of the self-published books in the world as of 2019 (\cite{curcic2023amazon}); in practice, self-published writers are subject to whatever Amazon decides for its business model. The advantages of self-publishing are that a writer retains full creative control and receives a larger share of royalties than what a publishing company would typically offer. The disadvantages are that the writer needs to do more work themselves (or pay more people themselves) to turn the manuscript into an actual book, and the writer also lacks the reach of a traditional publisher which is able to make deals with large distributors and to undertake ambitious advertising campaigns. Typically, a self-published writer advertises through Amazon itself and through social media.

A third model that needs to be mentioned here is the market for short stories, which functions somewhere in between the extremes of traditional publishing and self-publishing. Typically, writers do not rely on a literary agent to sell short stories, but instead submit stories individually to the editors of fiction magazines. These magazines are open to submissions either continuously or during certain periods of the year. Editors review the submissions and decide which ones are best to publish in the next issue of the magazine. If a writer's story is selected, they sign a contract selling rights to the magazine and receive a fee. These fees are usually fairly low - with ``professional rates'' in science fiction, for instance, consisting of about eight cents per word for a story of a few thousand words (\cite{tolbert2019sfwa}). Anthologies (books of short stories from a mixed group of writers) work in a similar way.

According to the queer and disabled writers in our study, each of these branches of the fiction-writing industry is precarious. In traditional publishing, the primary issue is market consolidation. The number of available publishers is shrinking, and consequently, so is the number of traditionally published books that can be sold and the level of artistic risks publishers are perceived as undertaking. Furthermore, in a consolidated market, the majority of the available money goes to a smaller and smaller number of writers; in the Authors' Licensing and Collecting Society's survey, 24\% of writing income was distributed to just the top 1\% of writers (\cite{authors2022why}).

\begin{quote}
    Generative AI is one element of the lengths corporations are going to to consolidate power and franchise. Both are alarming, but one is the cause. \textit{(Writer C)}
\end{quote}

Editors at these publishers tend to be understaffed, overworked, and consequently neglectful. The budget available for marketing a book is also increasingly small or unavailable, and little data is available to writers regarding their books' actual sales and reception. As a result, the seemingly impressive milestone of having traditionally published a novel can result in very little real sales or income.

\begin{quote}
    Everyone I know is complaining about publishers expecting authors to do all of their own marketing themselves, despite this being one of the main reasons to deal with a publisher. The common complaint being that in order to be an author, you're expected to be a social media influencer as well. \textit{(Writer R)}
\end{quote}

\begin{quote}
    Social media makes it easier to get superficial attention, certainly. I really can't tell most of the time what the true level of engagement with my work is. Even if it gets into a book or lit mag, is it being read? Who knows. \textit{(Writer D)}
\end{quote}

While publishers typically provide little support to an writer in creating and selling their book, writers also perceive them as fickle, preferring sure hits over books that appeal to a niche market or that take artistic risks, and dropping writers quickly if a book does not perform to expectations. This creates a double bind where traditionally published writers do not feel supported in selling their books, but are punished when the books do not sell.

\begin{quote}
     Building a readership on twitter is more difficult, also trad pub seems to want to bet on established writers more often than newer writers who don't write to current trends. \textit{(Writer T)}
\end{quote}

\begin{quote}
    My most recent series also launched during lockdown when bookshops were shut, which almost certainly had an effect on sales, and therefore makes publishers less interested in future projects. \textit{(Writer J)}
\end{quote}

\begin{quote}
    I expect that I will have a book that flops. The question is how willing will publishing be to trust that the book after the flop will hit. Authors, brilliant authors, get abandoned if they waver in profit. Which is why I keep my eye on other models, so I can pivot fast if I hit a snag. \textit{(Writer K)}
\end{quote}

Advances for a traditionally published book are determined individually in negotiation with the writer's agent (there is no ``standard advance'') but are typically quite low, as are other payments writers typically expect, such as the fee for writing a short story. The Author's Guild income survey describes the median income of a full-time writer as below minimum wage (\cite{authors2023key}). While some degree of economic struggle has always been inherent to the creative arts, this dearth of income is not simply business as usual; according to the Authors' Licensing and Collecting Society, median income for full-time professional writers has \textit{declined} by 60\% between 2006 and 2022. The same report shows that gender pay gaps have increased during this time and that the number of writers able to write full-time has fallen dramatically (\cite{authors2022why}). The marginalized writers in our survey are keenly aware of this drop in earning ability and are worried about their futures.

In self-publishing, this set of problems is replaced with a different one. Rather than competing for a limited number of spots with a traditional publisher, self-published writers compete directly for reader attention with an unlimited number of other self-published writers. On some platforms, such as Kindle Unlimited, a fixed pool of money is distributed between all writers depending on their number of page views, turning the competition for readers into a zero-sum game. Thus, the primary issue is how to stand out and bring one's work to readers' attention in a crowded field. There is pressure for a self-published writer to write very quickly and to appeal to readers who can be equally fickle as traditional publishers.

New platforms like Patreon and Substack constantly appear, offering alternative business models that promise to help writers find stability, but in practice these platforms quickly deteriorate in quality, changing their business models once they have captured a user base. In particular, the social network X (formerly Twitter) was until recently one of the most reliable places for writers to connect with readers and promote their work, but changes in its algorithm and user base in the past several years have rendered it largely useless for this purpose.

\begin{quote}
I do think that making money will become harder for authors in general. The demise of twitter has been a large blow to independent publishers. It was one of their main promotional avenues. With the changes to that platform many users have left and it is much more difficult to promote books and work on the site. Amazon getting rid of its magazine subscription service also dealt a blow to magazine subscription revenue. A magazine like The Dark Magazine, had to drop it's [sic] per word rate from 6 cents down to 5 and is now publishing reprints where before they only published originals \dots I expect private equity to take over a lot of the resources we use and this will make it harder to gain readers attention because tools that were once free will become subscription based. Pay for visibility style social media seems to be the way everything is going.  \textit{(Writer T)}
\end{quote}

Events like the COVID-19 pandemic, prior to the rise of genAI, exacerbated this instability for all sectors.

\begin{quote}
    Since the pandemic it's become much more difficult for writers to get readers' attention. For a long time there were fewer in person opportunities to meet readers. Virtual conventions and events were all we had and while that was great for accessibility those opportunities have diminished in the last year. \textit{(Writer T)}
\end{quote}

\begin{quote}
    The 2020 quarantine resulted in a lot of new novelists on the market, and consequentially more authors are drowned out. \textit{(Writer S)}
\end{quote}

Because writers ``write for love'' and because many non-writers dream of writing books of their own, there is a sense of vocational awe (\cite{ettarh2018vocational}). As with vocational awe in other sectors, this comes with the assumption that writers do not require benefit such as living wages or job security, nor is there any real material incentive for publishers to provide these things. As a result of all these factors, even very successful queer and disabled writers expressed uncertainty about the future of their careers.

\section{Effects of GenAI}

\subsection{Effects on the Industry}

Table \ref{tab:before-after} shows a summary of how genAI has exacerbated the precarity that fiction writers experience. While this precarity is not solely caused by AI, genAI exacerbates almost every aspect of the problem.

The most severe effect of genAI, and the one our writers discussed most, is the large influx of AI-generated books into self-publishing spaces and short-story markets. As discussed, the large number of writers competing for reader attention was already a challenge in self-publishing, but the influx of genAI has made it dramatically worse. In short story publishing, editors have found it challenging to cope. In most cases, only a single editor or small team manages submissions for a given magazine and submissions are managed by hand, which means even a relatively small influx (say, a few hundred submissions) can pose challenges. Most famously, Clarkesworld, one of the most venerable and highest-paying online science fiction magazines, temporarily closed to submissions in early 2023 (\cite{clarke2023written}). While Clarkesworld eventually improved its procedures and reopened, AI-generated submissions continue to be a source of stress for Clarkesworld's editors a year and a half later, persisting longer than the other problems the magazine faced in 2023 (\cite{clarke2024status}). Other well-known magazines have either cited AI as a reason for closing (\cite{dawson2024closure}) or continue to struggle with waves of up to 1000 low quality AI-generated submissions during open periods (\cite{tobler2024deadlands,uncanny2024please}).

\begin{quote}
    Generative AI is already flooding the self-pub market, making it very hard for readers to discern what is real fiction and what is machine-generated trash \dots
Unfortunately I don't know what can be done on the self-pub level, and I suspect that authors without existing name recognition, or without a recognised publisher behind them, will struggle to get found amidst the swamp \dots
Self-pubbed and indie-pubbed authors are probably going to struggle, since there will undoubtedly be many, many more "books" published at that sort of level. Readers will either have a far harder time finding decent books amidst the AI dross, or will give up completely on trying to find anything that doesn't come from a traditional publisher with a recognised policy against using AI "authors". \textit{(Writer J)}
\end{quote}

In context, this is not a matter of human writers being out-competed by the equally good and prolific work of genAI models. Rather, editors have reported that the fiction that these models produce would not have been published on its own merits ("among the worst submissions we've ever received," according to Clarkesworld (\cite{clarke2023written})); the problem is that, due to the sheer volume, the editor's inbox becomes flooded with what is essentially spam. Proposed solutions to this problem, such as using AI detection software or blocking particular IP addresses, are problematic because of the human writers who would also be blocked in this process. For example, most of Clarkesworld's AI submissions were sent from particular countries in the Global South, but many already-marginalized and underrepresented human writers come from these countries as well (\cite{clarke2023written}).

In traditional publishing, writers can typically only submit with the help of an agent. However, traditionally published writers worry that their publishers will put AI-generated cover art on their books without their consent, that publishers will sell their books to tech companies as training data without their consent, or that publishers will use AI to evaluate submitted manuscripts instead of having a human read them. (The midsized science fiction publisher Angry Robot Books recently proposed doing the latter during an open submission call (\cite{portsmouth2024storywise}), but backtracked less than a week later after significant social media backlash (\cite{coker2024open})).

\begin{quote}
    I think trad publishers are going to try to work generative AI into all of the processes in an effort to reduce costs. Many trad publishers are requiring that you sign an AI generated release in book contracts. These releases are non negotiable. Angry Robot announced recently that they wanted to use an AI service to help read, rate and sort their submission slush pile. They reversed the decision for this submission round due to push back from authors but did not comment on future submission periods. I believe that many publishing houses will look at implementing these systems without telling authors. These type of programs are detrimental to authors who write ""outside the box."" AI programs are only able to rate and grade work using the data scraped from other authors work. The biases of the creators of the program will lead to these programs. \textit{(Writer T)}
\end{quote}

To our surprise, most of the queer disabled writers in the study did not specifically mention their status as queer and disabled as a reason for being more pessimistic about the future. For example, although there is evidence that genAI has a heteronormative bias and struggles to generate stories about queer characters (e.g., the story about lesbians in \citet{ippolito2022creative},) our writers, apart from Writer T above, did not generally point out that AI's outputs will be biased against them. Instead they are more likely to express pessimism about genAI and about the future in general, universal terms. However, as the quantitative results in Tables \ref{tab:demographic}, \ref{tab:queerquant}, and \ref{tab:straightquant} show, queer disabled writers are in fact hit harder by this sense of general pessimism. One writer who did obliquely mention anti-queer sentiments is Writer R, who writes furry fiction - a niche genre involving anthropomorphic animal characters, which is heavily queer and heavily stigmatized.

\begin{quote}
    A gallery site I use to host my writing alerted its users that they had reason to believe everyone's work had been scraped. I have only a very limited reach to begin with, and it's to a community that vehemently rejects so-called AI, so the only practical effect was making me angry that my work was stolen. Especially since the people stealing that work have for years argued that said work shouldn't be allowed to exist and had no value. \textit{(Writer R)}
\end{quote}

\subsection{Views of AI}

As can be seen in Table \ref{tab:queerquant}, queer disabled writers are remarkably uniform in their pessimism about genAI. These marginalized writers' views of genAI are characterized by two major factors. First, due to the central role of intentionality and self-expression in their writing, they do not view AI as useful for this task. Second, due primarily to the use of pirated training data in genAI, they view the use of these AI tools as unethical.

\subsubsection{Ethical Concerns}

It is well known that LLMs are trained, in part, on Common Crawl data scraped from the Internet, including many works of copyrighted fiction posted online, and that these copyrighted texts can in some cases be memorized and repeated by the model verbatim (\cite{carlini2021extracting}). These models are also trained on datasets of pirated books, such as Books3, for which the writers were never compensated (\cite{reisner2023revealed}). While self-published writers can put notices inside their books saying that they do not consent to use of the book as training data (\cite{authors2023practical}), AI companies tend to scrape content indiscriminately and are not likely to respect these notices. (Presumably, the authors did not consent to their books being pirated, either.) Meanwhile, in at least some cases, the models trained on this data are in fact used to create``knock-offs'' or recognizably plagiarized versions of human writers' work, which are self-published and put into digital marketplaces to directly compete with the original writer (\cite{tapper2023authors}).

At the time of writing, numerous class-action lawsuits have been filed against AI companies by groups of fiction writers, nonfiction writers, journalists, visual artists, musicians, and even open-source software developers, all centered around this problem of copyrighted work being used without permission as training data (see \citet{baker2024case} for a summary of current cases,) The queer disabled writers in our study are uniformly on the side of the plaintiffs in these lawsuits and consider the use of any LLM, for any purpose, to be plagiarism.

\begin{quote}
Copyright exists for a reason and if you want to use copyrighted works, you have to pay for them. Speaking as an author AND an AI researcher, most AI researchers - even those who understand how the systems work at a technical level - do not understand the function these systems are computing. Language models have the effective capacity to recreate their inputs and therefore are effectively functioning as plagiarism engines when trained over copyrighted text.
Money should flow from reader through publisher to author in exchange for the author and publisher providing the reader something to read. Plagiarism subverts that by cutting the original author and publisher out of the loop so they cannot recoup the costs of producing their content. \textit{(Writer N)}
\end{quote}

The typical defense to these accusations of plagiarism is to cite the U.S. doctrine ``fair use'' and to view the use of copyrighted training data as something more akin to sampling and remix culture in music (\cite{lemley2020fair}). However, we did not find any queer disabled writers in our data who view it this way or who are excited about the possibilities of having their or others' writing remixed by LLMs. It remains to be seen whether any writers in our control group take this view.

Many writers also cited other ethical concerns, such as the energy and climate cost of genAI, the proliferation of misinformation and scams, and the disruption of public online services such as news and search. These concerns are important but are not unique to fiction and have been extensively discussed elsewhere (e.g., \citet{bender2021dangers, jiang2023ai})

\subsubsection{"Soullessness"}

As has been mentioned, fiction writers ``write for love,'' which has three components: an emotional attachment to the activity of writing, a sense of purpose in communicating something through writing, and a desire to connect meaningfully with a reader. Intentionality and a conscious wish to communicate are inherent in the process. Moreover - as the writers in our study repeatedly explained - these traits are central to what makes a work of fiction valuable in a writer's eyes.

We will refer to this concept of value derived from conscious, intentional communication as ``soul.'' By using this word, we do not mean to make any ontological claims about actual souls, but simply to adopt as an in situ concept a term which the writers themselves used with surprising frequency.

To have \emph{soul}, in its most basic manifestation, means that a writer consciously wishes to use fiction to communicate something they think and feel. They make an intentional effort to translate these thoughts and feelings onto the page into a form to which a reader can appropriately respond. To a writer with this kind of soul, each work of fiction has conscious, intentional meaning and this meaning is meant to be transmitted, even if it will be received differently by different readers.

\begin{quote}
    Writing is thinking, and art is about humans making things special and sharing them. Much of the joy of writing and art is about humans trying to understand their worlds, learning what's special about them, and deciding to share their discoveries. While there's great value in generative AI, it is only replicating a small part of the human experience, and we should make sure that it does not disrupt the cycle of understanding, making special, and sharing that makes us human. \textit{(Writer N)}
\end{quote}

By contrast, a genAI's writing is viewed as ``soulless,'' coming from a process which is seen as automatic or mechanical, and therefore - crucially - devoid of any intentionality or genuine meaning. Rather than creating art, even bad or unskilled art, the AI is seen as producing something fundamentally counterfeit, which lacks art's most central and crucial component.

\begin{quote}
    A machine can NEVER do what we do. We pour our hearts into the things that we write, and learning that that work has been stolen to feed a machine that regurgitates words based on formulas is soul-crushing. \textit{(Writer A)}
\end{quote}

This concept of ``soul'' bears a close resemblance to the argument made by \citet{jiang2023ai} that, in philosophies of art from around the world, art is a specifically human process in which there is conscious intent to communicate an individual's culturally situated affective experience to an audience. According this definition, the activities of an image generator, which does not have a grounded understanding of the experience it is conveying or the perspective of its audience, cannot be described as art. \citet{jiang2023ai} go on to describe several technical reasons why the inner processes of art generators are not similar to the processes of human artists; these, as well as other philosophical takes on intentionality in genAI such as \citet{redaelli2024intentionality} or \citet{guckelsberger2017addressing} are beyond the scope of our paper.

The idea of co-creativity, or of worthwhile art which could result emergently from the interaction of a ``souled'' human with a ``soulless'' device, is one that the writers in our study do not appear to be aware of. This is in marked contrast to studies of genAI use in other fields. Many of the artists in \citet{kaila2024gardening}, for instance, embrace the idea of being less personally involved in the work, becoming ``gardeners'' or ``curators'' of the AI's work rather than direct creators. Perhaps because of the relatively solitary and individual character of writing, and perhaps because they are less likely to see their own experiences reflected in ``mainstream'' literature, queer disabled writers do not want to become gardeners. Several expressed that even if genAI did allow them to create books faster, and without significant ethical issues, they would still not want to use it due to its lack of a soul.

\begin{quote}
    It cannot do any part of my job better than I can do it myself. And even if it could, I would not buy a machine that would eat my ice cream for me. \textit{(Writer K)}
\end{quote}

What we have described above is the less intense variation of the concept of ``soul.'' Some writers espoused a more intense version. The more intense versions of ``soul'' involve difficult and intentional effort; deep emotion; and the communication of intensely personal feelings and experiences. Although the work being created is fiction, a writer can ``pour their heart out'' into the work in a way that reflects very strong emotions, deeply personal life experiences, and even personal traumas. It is consequently even more of an affront to these writers for their highly personal work to be replaced by a ``soulless'' AI-generated work.

\begin{quote}
Everything I write--and I write fantasy, which is probably the most fictional of all the fictions-is a product of my life and what I experienced. Every character, every setting, every dramatic situation, every theme, every coined word or sentence level decision is a result of the entirety of my life and my experience. It is an illustration of how my life has made me, how the world and my time in it has touched me, from the gentle cup of my mother's hand on the back of my skull as she lifted me from my crib to the hands around my throat while my ex boyfriend screamed \dots It's about everything I ever read, everything I ever learned, everything I ever wished could happen or would happen, everything I'm furious about, and everything that left me terrified that it might. 
Novels, which is what I primarily write, are stories that are entertaining (we hope) but are also powerful devices for empathy, understanding, realization, and changes of heart (we also hope.)  \textit{(Writer K)}
\end{quote}

In short, to the queer disabled writers in our study, genAI is useless, because it cannot replicate the most important aspect of the writer's process; and it is worse than useless, because in the process of its attempts at writing, it causes various ethical harms.

As a result of these attitudes towards genAI, most queer disabled writers in our data had never tried to use genAI, and most stated that they would not do so under any circumstances. However, a few writers did describe experiences where they experimented with genAI. The initial experience with genAI is characterized, not necessarily by desire to use the AI in published work, but by simple curiosity about its abilities. For some of the writers in our data, this curiosity led to disappointment with output which, to the writer's expert eyes, was not high in quality, and the writer subsequently stopped using the AI.

\begin{quote}
I played around with using ChatGPT when it came out to see what it could do with poetry, dialogue, various writing things. I was disappointed in the results. I am told ChatGPT has improved, though it's poetry doesn't seem to have improved... \textit{(Writer D)}
\end{quote}

For other writers with more technical expertise, the AI did prove useful in a limited capacity, as a brainstorming tool, as a tool for preliminary research, or a producer of raw material that the writer could then modify and combine with other inspirations as they saw fit.

\begin{quote}
I experimented with generative AI to create a movie script, but abandoned it. I have used it to create illustrations \dots but the ability of the systems to take art direction was poor. 
\newline
I have used it to create marketing materials for books, but while the content of the suggestions was good, the actual text was unusable. 
\newline
I do not use generative AI in the writing process for my own fiction series as I want the inspirations for and the text of my own series work to be either wholly my own or my own with advice from my human editors, adaptation writers, and translators \dots
\newline
    I have used Midjourney, ChatGPT+DALL-E, Stable Diffusion, Adobe's generative tools, upscalers, and several other visual models, principally for ideation but sometimes for generation of elements when the client specifically wanted an AI look. Generally these systems cannot take art direction and are not useful for producing final product, but can be helpful to a Photoshop jockey. \textit{(Writer N)}
\end{quote}

Only three queer disabled writers (D, M, and N) indicated that they have ever used genAI for writing or related tasks. Due to our process of theoretical sampling, we have already analyzed all three of these writers' responses. Because the number is so small, any firm conclusions about fiction writers' direct experiences using genAI will need to wait until we have also analyzed the non-queer/non-disabled control group.

\subsection{Mental Health Effects}

Because genAI has deepened the precarity of an already very precarious industry, and because of our writers' ethical opposition to AI (described above), many queer and disabled writers have experienced strong emotional reactions to the use of AI. Most strikingly, for this specific population of writers, a large number said that the rise of genAI had been harmful to their mental health. This largely takes the form of an exacerbation of existing mental health difficulties. While the specific difficulties can vary - various writers described anxiety, stress, resentment, anger, sadness, depression, and hypervigilance, with few listing more than one of these - almost all of these writers told the same underlying story, which is that it was already difficult to maintain good mental health as a disabled writer in the publishing industry, and that the rise of genAI has made this difficulty notably worse.

\begin{quote}
    I was already somewhat stressed and anxious about the prospect of building myself a working career as a writer, and the rise of generative AI has simply magnified that anxiety. It is becoming progressively more difficult to imagine myself being truly successful as a writer and being able to rely on writing as my primary source of income, which I would really love to do, as I am disabled and struggle to find other work. \textit{(Writer A)}
\end{quote}

In addition, the ubiquity of genAI - being advertised widely, added to many existing forms of software, and constantly discussed - makes it hard to escape from these concerns. Coping strategies which would normally help when there is an ongoing problem in a certain area, such as making time to disconnect from that area and focus on something else, become more difficult when the problem is hypervisible and seems to be everywhere at once.

\begin{quote}
    there's no escape now. it's in every browser, every search engine, it's in Word, which I have to use, it's in my google docs, which I no longer use for betat [sic] reading and critique. And It's not like I can simply print my manuscript and do everything on paper from here on in, that's ridiculous. \textit{(Writer K)}
\end{quote}

Even when not prompted to describe their emotions or mental health, several of the writers responded to our questions with strong emotion and marked animosity, to the point of making apocalyptic predictions and wishing harm on AI researchers.

\begin{quote}
    I hate "AI" as I hate Hell and all montagues\footnote{``Hell and all Montagues'' is a reference to ``Romeo and Juliet.''}. I will resist it with all my might. if we don't, art is dead, and humanity soon after. \textit{(Writer K)}
\end{quote}

\begin{quote}
    Also if you're attempting to "improve" generative AI then I hope your computer always crashes with your work unsaved, from now until the sun swallows the Earth. \textit{(Writer J)}
\end{quote}

These impacts on mental health should give us pause. Although there is a stereotype of mental illness being associated with artistic talent - and although many creative professionals are in fact mentally ill - the truth is that untreated mental illness makes most activities, including the production of creative output, more difficult. A widespread increase in the severity of mental illness among creative writers who were already ill means a high likelihood of some of these writers losing their ability to continue to write - and therefore, of the unique viewpoints of this already marginalized population being silenced or marginalized further.

\section{Strategies of Response}

Given that the fiction writing industry is inherently precarious, and that genAI has worsened this precarity, it is not surprising that fiction writers use strategies to protect themselves, both individually and collectively. On an individual level, writers pursue the quality of ``momentum'' which allows them to increase their resilience by building on past success, and writers also set personal boundaries, even at perceived personal risk, to avoid being involved with the use of genAI. On a collective level, despite being blocked from some traditional means of collective action, writers have organized both formally and informally to create situations of legal and social pressure which discourage the use of genAI in their industry.

\subsection{Momentum: Building Personal Success}

Momentum is the tendency for previous successes, in a writing career, to make future successes easier. The more that a writer's successful work is recognized by publishers and readers, the easier it will be for the writer to sell their work to those publishers and readers in the future. Momentum not only includes broad measures like sales numbers, but also the idea of building a ``niche'' or a ``brand'' - a particular style of writing or set of themes that readers know to expect from a particular writer. Experienced writers throughout our data describe the importance of momentum, and the ways that their own momentum has benefited them. In fact, momentum was the only individual protective factor against any form of precarity that writers mentioned in any of the analyzed data.

\begin{quote}
    It's become easier for me because I've built a bit of a brand \dots and it's been helpful in promoting my work to readers. Also, because of my previous sales, community work and award nominations I've had a bit of an easier time convincing people to try my work. \textit{(Writer T)}
\end{quote}

\begin{quote}
    I'm lucky enough to myself be, technically, a niche publisher in a very devoted subculture, which lets me network to other authors and publishers. But "go become a furry" isn't practicable advice for most authors. \textit{(Writer R)}
\end{quote}

Inexperienced writers are also very aware of the importance of momentum and actively plan to create momentum for themselves.

\begin{quote}
    I am hoping that making money as an author will become easier for me over the next five years, because I hope to publish my first novels, along with more short stories. I also hope to start building a fan base that will recognize my name and more consistently purchase my work. \textit{(Writer A)}
\end{quote}

However, momentum is not absolute; once it is established, it can still be stalled or stymied for many reasons. Even writers with a great deal of momentum - in the form of successful books and even award nominations - cautioned us that their situation is still, in many ways, not secure.

\begin{quote}
I can't deny that it's easier for me now. I have the evidence of book sales - my first statement sales numbers are higher with each ""tentpole"" book I've released \dots The trick is keeping that curve, though, and I don't trust that I can. I have to get readers to trust me all over again every time I release a title, because while I am generally a fantasy writer with strong romance threads with protagonists of all kinds of sexualites who bangs on about social justice in very obvious ways, there's always something different. \textit{(Writer K)}
\end{quote}

\subsection{GenAI Boundaries}

Many writers in our data also have attempted to act against genAI through personal boundaries. Rather than safeguarding the writer's own career, genAI boundaries are set in solidary with other creative humans whose careers are at perceived risk from the use of genAI, even if this increases the cost to the individual writer. They also protect the writer from the moral risk of becoming directly involved with a technology that they find unethical and distasteful and that may harm them.

In our data, we have found four distinct levels of genAI boundaries, some more severe than others. At the first and mildest level, writers seek to prevent AI from being used in relation to their own work. This is challenging for traditionally published writers who do not always have direct control over how the publisher will go about creating their book's cover, audio version, translations, or advertising materials. One way of trying to gain this control is by negotiating for it in the book's contract; however, this strategy is not without risk.

\begin{quote}
    I want terms that all of my production and promotion be done by people who don't use AI in their work. I don't want anything to do with it. And that might mean that my career in traditional publishing is over, no matter how bright my future looked. \textit{(Writer K)}
\end{quote}

An experienced literary agent, who we contacted for more context about this issue, confirmed that these types of AI clauses are a point of contention in the majority of current book contracts in her genre. Some publishers are friendly to these clauses, while others will not consider them. Sometimes even when a publisher will not promise to avoid AI, a writer can still indirectly prevent some uses of AI - for example, by asking for approval over potential narrators (\cite{bowman2024personal}).

At the second level of genAI boundaries, writers refuse to work with publishers that are known to have used AI (e.g., for editing, translation, or cover art), regardless of whether AI will be used on that writer's work in particular.

\begin{quote}
    I've had to start being more watchful \dots you always have to keep an eye out for if publishers are using AI covers, etc, and so be aware of who you don't want to publish with (making things harder, given how difficult it is to get published anyway). \textit{(Writer J)}
\end{quote}

At the third level, writers refuse to use software that contains AI features, such as word processors, search engines, and social media - even when this entails a significant increase in the difficulty of the writer's everyday labor.

\begin{quote}
The need to avoid it has added so much work, and so many extra steps, to my work. I can't use programs I used to rely on, or social media platforms that used to promote my work, and tools I still do use because I have no alternatives require continual opting-out \dots \textit{(Writer R)}
\end{quote}

At the fourth and most severe level, some writers set genAI boundaries by cutting professional and social ties with anyone who is not also boycotting genAI. (The writer who discussed this did not consent to being directly quoted.)

When we have analyzed all the data, we will be able to estimate the relative prevalence of these various levels; for now, what we know is that all four levels exist. It is still possible we may also discover additional levels.

None of these boundary-setting strategies are risk-free for writers and some have obvious costs and harms. Writers engage in these strategies anyway due to the strength of their convictions against genAI and their solidarity with other human artists.

\subsection{Public Backlash and "Speaking Out"}

The third strategy used by queer and disabled fiction writers to resist genAI is public and collective backlash. Groups of creative workers who have the right to unionize, such as the WGA, have undergone collective action against AI through their unions with some success (\cite{anguiano2023hollywood}). Prose fiction writers engage in their own forms of collective action - but, because they are freelancers who cannot unionize and collectively bargain, this collective action instead takes the form of decentralized social media discourse and social pressure designed to discourage any use or acceptance of genAI in publishing. Writers in our data discussed public discourse, not merely as venting or making their viewpoints clear, but explicitly as a collective means of limiting AI-related harms.

\begin{quote}
    I hold hope that voices for human creation can remain strong and drive cultural change. \textit{(Writer C)}
\end{quote}

\begin{quote}
Generative AI is causing negative effects already, and I have no doubt it will continue to do so for a while. However, I am hopeful that the public backlash over it (see Angry Robot's recent open subs controversy as one example) will ensure that no reputable publisher will start using it in any real sense, especially since it's pretty garbage. \textit{(Writer J)}
\end{quote}

We briefly discussed the Angry Robot controversy above; social media backlash against the publisher, in that case, was strong enough that the publisher retracted their proposed use of genAI less than a week after it was first announced. In at least three cases in the past two years, traditional publishers - specifically, Tor (\cite{codega2022tor,stewart2024tor}) and Bloomsbury (\cite{weatherbed2023not}) - have faced similar waves of online backlash after using AI-generated art assets as part of their cover designs. As \citet{stewart2024tor} describes, this backlash was intentionally driven by BookTok influencers with the explicit goal of preventing AI from ``infiltrat[ing] creative spaces.''

Although the publishers did not retract the covers, Tor in both cases produced public statements saying it hadn't been aware the assets were generated using AI (\cite{codega2022tor,stewart2024tor}). For now, at least, publicly admitting to the \textit{intentional} use of AI is still too much of a PR risk for traditional publishers. Both these cases show that online backlash can in fact have an effect on publishers' choices and cause them to reduce, remove, or hide their intended uses of genAI. According to the literary agent we spoke to, traditional speculative fiction publishers are not yet using much AI in their workflows - in part because of the scandal it caused on the few occasions when they did (\cite{bowman2024personal}).

This is not to say that formal groups of writers do not also play a role. Although they cannot directly engage in collective bargaining, writers' groups like The Author's Guild have engaged in data gathering and advocacy geared towards informing policymakers about writers' values and needs related to AI (\cite{authors2023survey, international2023artificial}) and have joined in several of the ongoing class-action lawsuits (\cite{baker2024case}). However, the writers in our study seem overall less aware of these efforts and more aware of the informal efforts that are directly visible to them on social media.

Various scholars in the social sciences (e.g., \citet{mesler2024association}) have investigated these types of online call-outs, pile-ons, and ``cancelling,'' primarily in terms of the importance of these behaviors to a participant's political or in-group identity. However, this is an incomplete understanding. To cancel someone or something, as in the Angry Robot and Tor examples, has a material effect. Since marginalized writers view genAI as a source of increased precarity and of of harm to their mental health, a strategy of preventing the use of genAI is both individually and collectively self-protective. Lacking direct power over publishers, fiction writers opposed to genAI are still actively engaged in social strategies designed for their collective protection. Online vitriol against AI should therefore be viewed, not as mere ``polarization,'' identity formation, or virtue signaling, but as intentional strategies which serve to protect writers' material interests.

\subsubsection{Prescriptions}

Largely not being computer scientists or legal experts, the queer and disabled writers in our data so far have not suggested many specific policy improvements. The one theme on which they are nearly unanimous is that genAI should not be used to write fiction. Writers typically engaged in a rhetorical strategy we refer to as ``walling off the AI'' - insisting that while AI may be used for some other thing, it must not be used for art. Each writer has their own idea of what constitutes an actually appropriate use of AI - ranging from ``curing cancer'' to ``analyzing patterns in language'' to ``nothing.'' (Some of these suggested uses, such as curing cancer, are themselves tasks which require creativity - but which, when done by humans, do not centrally involve the expression of a human's ``soul.'')

\begin{quote}
    I believe that AI should be reserved for cases when both:
\newline - No humans want to do the task.
\newline - It is responsible to take the task out of human judgement.
\textit{(Writer C)}
\end{quote}

\begin{quote}
    I would like to express that it doesn't have to be like this. You don't have to write tools for scammers and spambots, that serve no function except to accelerate fake news stories, frustrate ordinary people, and prey on seniors, the desperate, and disabled. You could instead be using this exact same knowledge to explore the nature of human languages and grammars, enriching the entire world. \textit{(Writer S)}
\end{quote}

Our own tentative prescriptions, based on our own expertise and the knowledge we have gained from our data, are contained in the next section.

\section{Discussion}

A longstanding problem in the development of genAI is that tech companies eagerly promote the use of this AI in every industry, and breathlessly predict that every industry will be disrupted (in a presumably, overall positive way) but are actively uninterested in the specific material conditions in these industries, the actual needs of existing workers in these industries, or the ways that workers can be protected during these disruptions. The tasks necessary to understand and protect workers are explicitly or implicitly outsourced to other researchers with significantly less institutional power (e.g., \citet{eloundou2023gpts}.) Qualitative studies like ours are one way, albeit an inadequate way on their own, of performing this work and illustrating the actual situations that workers find themselves in as a result of technological change.

In particular, proponents of genAI often claim that it will ``democratize creativity'' since it can be used without special skills. This claim is made without real understanding of the existing conditions in specific creative industries. In the publishing industry as experienced by our marginalized writers, the biggest barriers to full democratic participation are not lack of skill or lack of technology, but precarity, uncertainty about the future, and low income. The presence of genAI so far appears to have increased these barriers rather than removing them.

Compared to other qualitative studies of creative workers' uses of genAI, the opinions of our queer disabled writers are unusually and overwhelmingly negative. There are several possible explanations for this. First, queer disabled writers' relative marginalization, results in a more severe experience of professional precarity, as seen in their strikingly lower income and lower chance of being able to write full time (Table \ref{tab:demographic}) - as well as a lower belief, as mentioned by Writer A above, in their ability to succeed in other industries. Although few of the writers mention their queerness or disability as a reason for pessimism, it is not surprising that a population which is already marginalized and experiencing worse outcomes would correspondingly be more pessimistic about outcomes in the future. We have yet to analyze the precise qualitative differences between these writers and the more optimistic ones in our control group - although the control group also broadly tends towards pessimism.

While this focus on queer disabled writers limits the generalizability of our results, it does not (in our opinion) diminish their importance. If a technology has disproportionately negative effects on a population which is already marginalized, this in itself is an ethical problem that should concern the technology's creators.

Second, the concept of ``soul,'' as describe above, is relevant to fiction writers due to the relatively solitary nature of their creative practice. While it is of course possible for multiple fiction writers to collaborate on a project, and while fiction writers do receive some feedback from beta readers, editors and sensitivity readers, the overwhelming majority of their creative work is done alone. Consequently, it may be much easier for a fiction writer to ``pour out their heart'' into a work, and to feel a strong sense of personal and emotional ownership over this work, compared to a worker in video games or industrial design who primarily works in a team-based setting. This is not to diminish the creativity of this latter group of workers or their ability to personally express themselves, but simply to point out a difference in how their creativity is channeled and expressed. Creative workers in industries where individual self-expression is paramount may have a harder time embracing genAI than creative workers in industries with a more team-based or functionality-based approach. However, even studies in fields such as industrial design have shown that concern does exist in these industries over ownership of AI output, over perceived lack of control over the adoption of the technology, and over the possibility of human workers being replaced by AI; this anxiety is highest among the lowest ranking workers with the least power (\cite{uusitalo2024clay}).

\subsection{Policy Implications}

The most obvious and unanimous policy view among the writers in our dataset is that copyrighted creative work should not be used as training data without consent and compensation to the human creators. We agree with this view and note that it is also the view taken by the plaintiffs in most of the current class action lawsuits against genAI companies (\cite{baker2024case}). As of this writing, the results of these lawsuits have yet to be decided, and the regulatory space around genAI is underdeveloped.

Based on the expressed experiences and needs of the queer disabled writers in our dataset, we have thoughts on several potential policy prescriptions, which we summarize below.

\begin{itemize}
    \item{First: ideally, either through regulation or through the success of one of the aforementioned lawsuits, the use of copyrighted creative work as AI training data without consent of, or compensation to, the creator should be recognized as plagiarism and disallowed.}
    \item{Transparency and disclosure about the origins of genAI-assisted work are crucial. Requiring AI-generated or AI-assisted work to be labeled as such would allow readers who value ``soul'' to find it and would prevent such readers from being fooled into buying AI-generated work; it would also help to prevent the spread of AI knock-offs of a specific author's work, as described by \citet{tapper2023authors}. It would also serve as no hindrance to readers who are genuinely interested in reading AI-generated work; in fact, it would help these readers as well by making it easier for them find such work.}
    \item{One way of mandating transparency about AI-generated work is through the use of watermarking - as described, for example, by \citet{kirchenbauer2023watermark}. We support the use of watermarking to help make AI-generated content more recognizable. However, even mandatory watermarking would not necessarily fix our writers' problems; as long as alternate models are available which do \textit{not} watermark their output, it will still be easy for unethical individuals to use these models to flood e-bookstore shelves and magazine submission inboxes with unwanted AI-generated work.}
    \item{Recently - several months after our data was collected - the publisher HarperCollins struck a deal with an unnamed AI company whereby authors of nonfiction could opt in to the use of their books as training data, in exchange for a fee of several thousand dollars to be split between the writer and publisher (\cite{albanese2024agents}). We are cautiously in favor of consensual arrangements like these, so long as they remain voluntary; the primary ethical objection of our writers to genAI is its use of books as training data without compensation or consent, so we support the creation of agreements which will introduce compensation and consent into the data-collection process. The HarperCollins deal's treatment of AI training rights as subsidiary rights, to be sold separately from print and ebook publication rights, is also consistent with how other subsidiary rights are treated in publishing, such as the right to produce audiobooks or translations. However, we suspect that the majority of writers we surveyed would \textit{not} opt in to an arrangement like this, even for a considerable fee - due to their disdain for computer programs that lack a ``soul,'' their strategy of protecting their own industry by disavowing all uses of AI, and their concerns about other ethical issues with genAI, such as disinformation and environmental costs. If consent to AI training is to be voluntary and meaningful, then the preferences of these writers with regards to the use of their own work must also be respected.}
    \item{Solutions have been proposed, such as the Shapley Royalty Share Framework (\cite{wang2024economic}), in which the contribution of various human writers to a genAI's output is calculated by some means and writers are compensated in proportion to their calculated contribution. Such frameworks are conceptually similar to schemes in many countries which compensate writers for certain other uses of their work, such as Canada's Public Lending Right, which compensates writers whose books are borrowed in libraries (\cite{arts2024public}), and Access Copyright, which compensates writers for university photocopying of their work. Notably, even though both of these uses of a writer's work are considered ``fair dealing'' in Canada (the equivalent of the American concept of ``fair use''), the government ensures writers are compensated for them anyway. A scheme of this nature to compensate writers for the use of their work by genAI would correct the problem of writers' work being used without compensation; however, unlike the subsidiary rights deal discussed in the point above, proposed royalty-share frameworks such as \citet{wang2024economic} do not necessarily ensure that writers provide consent before their writing is used. The majority of writers in our study would clearly prefer to deny consent for their writing to be used in the first place, since their professional and ethical objections to genAI go beyond the mere issue of compensation. The measures proposed by \citet{wang2024economic} and others are also expensive to calculate.}
    \item{If the regulations described above do not discourage technology companies from scraping training data from plagiarized books, another potential technique to stop this behaviour is the use of data poisoning (\cite{wan2023poisoning}), by which training examples can be deliberately designed cause harm or undesirable behavior in an AI system that trains on them. In visual art, tools have already been released such as Nightshade, which allows artists without AI or programming expertise to introduce small permutations into their art which will poison any AI model that trains on it against that artist's wishes (\cite{shan2024nightshade}). In an environment where few other means of control are available to creative professionals, it may very well be reasonable for them to create these incentives to force companies to take their wishes seriously.}
    \item{Several writers in our data express concern about particular scenarios that, to our knowledge, have not yet come to pass. For example, some worry that instead of accepting original work from human writers, major publishers will switch to a model of generating a book draft using AI and hiring a human to revise it - with less compensation than what would be provided to an original writer, and with significantly less freedom for the human writer involved to express their own ``soul.'' This is similar to one of the scenarios at issue in the recent WGA strike for screenwriters (\cite{anguiano2023hollywood}), where collective bargaining was used to prevent this scenario. Since prose fiction writers lack the right to collectively bargain, top-down regulation would be necessary for them to achieve the same end.}
    \item{Currently proposed antitrust regulation might indirectly help protect fiction writers, especially self-published fiction writers, from the worst harms associated with AI. In particular, self-published writers are dependent on Amazon due to its enormous market share, and Amazon has been unable or unwilling to weed out the majority of genAI-assisted content. If Amazon's monopoly were reduced - as advocated, for example, by the current head of the FCC in \citet{khan2016amazon} - then competition would be possible from online booksellers with stricter AI policies. In a genuinely competitive environment with many viable booksellers, writers, and readers with an interest in avoiding AI could gravitate to these booksellers, while AI-generated work could still be made and sold, by those with a genuine interest in the possibilities of genAI, at other stores with more permissive terms.}
\end{itemize}

Some of these proposed solutions are technological and could be implemented without regulatory action, but most would require change from the top down - either a high-profile success in one of the aforementioned lawsuits, or significant regulatory enforcement from applicable governments. While grassroots efforts from writers themselves are powerful enough to influence the decisions of a small company like Angry Robot, it is unlikely that grassroots efforts alone can change the practices of companies like Amazon, Google or Microsoft, which are among the most powerful in the world. Given the current political climate in the US, meaningful regulation and enforcement in the near future seem unlikely. However, we continue to urge lawmakers and others to consider the mental and material effects of technology on creative workers, especially the most marginalized creative workers, when making relevant policy decisions.

\subsection{Limitations}

We are aware that our results only reflect a particular set of fiction writers. As described above, we have decided to lean into some of our data set's unique demographic characteristics instead of trying to correct for them. Very little of our theory, except perhaps the results on mental health, is directly dependent on a writer's status as queer or disabled, and therefore it is quite possible the entire theory may generalize to the whole set of writers surveyed. However, it remains for our future work to verify whether this is the case. In particular, we have yet to analyze or account for the views of fiction writers in the control group who are more optimistic about genAI.

Although we have used the word ``marginalized'' - and although queer disabled writers are, by definition, marginalized - queer identity and disability status are only two among many possible axes of marginalization. While our survey materials asked about several forms of marginalization (race, immigration status, language, religion) we did not receive large enough numbers of responses in these categories to give them the intense focus that we were able to give to queerness and disability. In particular, both our data set and the general population of fiction writers are disproportionately white. We have attempted to mitigate some of these biases in our data through theoretical sampling and by deliberately seeking out BIPOC and otherwise marginalized voices within our data (some of whom, e.g., Writers K and T, are extensively quoted here.) However, this should not be construed as a substitute for a study directly focusing on BIPOC experiences in publishing, and more work needs to be done in this area.

Finally, because our data was collected mainly on Bluesky - and because Bluesky, along with other media outlets such as X and TikTok, is where a large amount of polarized discussion of AI takes place - it is possible that the responses we've received may overemphasize the importance of online discourse, or may be more polarized than a fully random sample of writers (queer and disabled or otherwise). Non-academic surveys such as \citet{authors2023survey} and \citet{humanity2023authors} also show widespread concern about plagiarism and widespread distrust of LLMs among writers, so we do not believe that the overall opinions expressed by writers in this study are minority opinions. Nevertheless, we do not attempt to claim that these opinions are the opinions of every writer - merely that they are opinions prevalent among queer and disabled writers which result from real concerns and which deserve our attention.

\subsection{Future Work}

As will be obvious to the reader and as we have described above, the first priority for our future work is to finish analyzing our data. If funding permits, we would like to additionally expand this study into a longditudinal form in which changes in writers' experiences and attitudes are tracked over time.

Although much ink has been spilled about the possible effects of current and future genAI on the arts, relatively little research has gone into the actual experiences of workers in the arts, especially those who are not already early genAI adopters and who are not the lead plaintiffs in the various ongoing lawsuits. We acknowledge existing studies which have taken this approach and we hope that further work engages seriously with creative workers' experiences in a wide variety of industries and contexts, particularly creative workers marginalized in various ways which were not fully included in the scope of this study.

As computational creativity researchers, we of course find AI's impact on the creative arts to be particularly interesting, but this does not imply that creative workers are unique or more worthy of protections than other workers. Any workers who are currently subject to precarity as a result of AI automation of work tasks, or to the imposition of genAI tools and processes from above, should be studied closely so that their experiences and concerns can be heard and addressed.

\section*{Acknowledgments}

This research was made possible by an internal grant from the Queen's University Fund for Scholarly Research and Creative Work and Professional Development (Adjuncts). We would like to thank Charlie Clarke, Kazjon Grace, Anna Jordanous, and Chris Liscio for insightful and helpful conversations.

\section{Conflict of Interest / Positionality Statement}

The first author of this paper is, by the definition used in this paper, a professional fiction writer (their fictional work has been published under a pen name.) Both the first and second authors identify as queer and disabled.

%Bibliography
\bibliographystyle{plainnat}  
\bibliography{ai-authors}

\begin{thebibliography}{46}
\providecommand{\natexlab}[1]{#1}
\providecommand{\url}[1]{\texttt{#1}}
\expandafter\ifx\csname urlstyle\endcsname\relax
  \providecommand{\doi}[1]{doi: #1}\else
  \providecommand{\doi}{doi: \begingroup \urlstyle{rm}\Url}\fi

\bibitem[Albanese et~al.(2024)Albanese, Milliot, and Maher]{albanese2024agents}
Andrew Albanese, Jim Milliot, and John Maher.
\newblock Agents, authors question harpercollins ai deal.
\newblock \emph{Publishers Weekly}, November 19 2024.

\bibitem[Anguiano and Beckett(2023)]{anguiano2023hollywood}
Dani Anguiano and Lois Beckett.
\newblock How {Hollywood} writers triumphed over {AI} - and why it matters.
\newblock \emph{The Guardian}, October 1 2023.

\bibitem[BakerHoestetler(2024)]{baker2024case}
BakerHoestetler.
\newblock Case tracker: Artificial intelligence, copyrights and class actions.
\newblock \url{https://www.bakerlaw.com/services/artificial-intelligence-ai/case-tracker-artificial-intelligence-copyrights-and-class-actions/}, accessed October 24, 2024, 2024.

\bibitem[Bender et~al.(2021)Bender, Gebru, McMillan-Major, and Shmitchell]{bender2021dangers}
Emily~M Bender, Timnit Gebru, Angelina McMillan-Major, and Shmargaret Shmitchell.
\newblock On the dangers of stochastic parrots: Can language models be too big?
\newblock In \emph{Proceedings of the 2021 ACM conference on fairness, accountability, and transparency}, pages 610--623, 2021.

\bibitem[Bowman(2024)]{bowman2024personal}
Hannah Bowman.
\newblock Personal communication, August 19 2024.

\bibitem[Carlini et~al.(2021)Carlini, Tramer, Wallace, Jagielski, Herbert-Voss, Lee, Roberts, Brown, Song, Erlingsson, et~al.]{carlini2021extracting}
Nicholas Carlini, Florian Tramer, Eric Wallace, Matthew Jagielski, Ariel Herbert-Voss, Katherine Lee, Adam Roberts, Tom Brown, Dawn Song, Ulfar Erlingsson, et~al.
\newblock Extracting training data from large language models.
\newblock In \emph{30th USENIX Security Symposium (USENIX Security 21)}, pages 2633--2650, 2021.

\bibitem[Clarke(2023)]{clarke2023written}
Neil Clarke.
\newblock Editor's {Desk:} {W}ritten by a human.
\newblock \url{https://clarkesworldmagazine.com/clarke_04_23/}, April 2023.

\bibitem[Clarke(2024)]{clarke2024status}
Neil Clarke.
\newblock Editor's {Desk:} {S}tatus report.
\newblock \url{https://clarkesworldmagazine.com/clarke_08_24/}, September 2024.

\bibitem[Codega(2022)]{codega2022tor}
Linda Codega.
\newblock Tor tried to hide {AI} art on a book cover, and it is a mess.
\newblock \emph{Gizmodo}, December 16 2022.

\bibitem[Coker(2024)]{coker2024open}
Desola Coker.
\newblock Open submissions 2024 update: 15th {April} 2024.
\newblock \url{https://angryrobotbooks.com/2024/04/open-submissions-2024-update-15th-april-2024/}, April 15 2024.

\bibitem[Colton et~al.(2016)Colton, Llano, Hepworth, Charnley, Gale, Baron, Pachet, Roy, Gerv{\'a}s, Collins, et~al.]{colton2016beyond}
Simon Colton, Maria~Teresa Llano, Rose Hepworth, John Charnley, Catherine~V Gale, Archie Baron, Fran{\c{c}}ois Pachet, Pierre Roy, Pablo Gerv{\'a}s, Nick Collins, et~al.
\newblock The {Beyond the Fence} musical and {Computer Says Show} documentary.
\newblock In \emph{Proceedings of the Seventh International Conference on Computational Creativity}, 2016.

\bibitem[Corbin and Strauss(2015)]{corbin2015basics}
Juliet Corbin and Anselm Strauss.
\newblock \emph{Basics of Qualitative Research: Techniques and Procedures for Developing Grounded Theory}.
\newblock SAGE, 2015.

\bibitem[Crenshaw(2013)]{crenshaw2013demarginalizing}
Kimberl{\'e} Crenshaw.
\newblock Demarginalizing the intersection of race and sex: A black feminist critique of antidiscrimination doctrine, feminist theory and antiracist politics.
\newblock In \emph{Feminist legal theories}, pages 23--51. Routledge, 2013.

\bibitem[Curcic(2023)]{curcic2023amazon}
Dimitrije Curcic.
\newblock Amazon publishing statistics.
\newblock \url{https://wordsrated.com/amazon-publishing-statistics/}, January 12 2023.

\bibitem[Dawson(2024)]{dawson2024closure}
Julie~Ann Dawson.
\newblock Closure announcement.
\newblock \url{https://www.bardsandsages.com/closure-announcement.html}, March 2024.

\bibitem[Eloundou et~al.(2023)Eloundou, Manning, Mishkin, and Rock]{eloundou2023gpts}
T.~Eloundou, S.~Manning, P.~Mishkin, and D.~Rock.
\newblock {GPTs are GPTs:} an early look at the labor market impact potential of large language models.
\newblock In \emph{arXiv preprint arXiv:2303.10130}, 2023.

\bibitem[Ettarh(2018)]{ettarh2018vocational}
Fobazi Ettarh.
\newblock Vocational awe and librarianship: The lies we tell ourselves.
\newblock \emph{In the Library With the Lead Pipe}, January 10 2018.

\bibitem[for~the Arts(2024)]{arts2024public}
Canada~Council for~the Arts.
\newblock The public lending right {(PLR)} program.
\newblock \url{https://publiclendingright.ca/}, 2024.

\bibitem[Guckelsberger et~al.(2017)Guckelsberger, Salge, and Colton]{guckelsberger2017addressing}
Christian Guckelsberger, Christoph Salge, and Simon Colton.
\newblock Addressing the “why?” in computational creativity: A non-anthropocentric, minimal model of intentional creative agency.
\newblock In \emph{International Conference on Computational Creativity 2017}. Association for Computational Creativity (ACC), 2017.

\bibitem[Guild(2023{\natexlab{a}})]{authors2023key}
The~Authors Guild.
\newblock Key takeaways from the {Authors Guild’s} 2023 author income survey.
\newblock \url{https://authorsguild.org/news/key-takeaways-from-2023-author-income-survey/}, September 27 2023{\natexlab{a}}.

\bibitem[Guild(2023{\natexlab{b}})]{authors2023practical}
The~Authors Guild.
\newblock Practical tips for authors to protect their works from ai use.
\newblock \url{https://authorsguild.org/news/practical-tips-for-authors-to-protect-against-ai-use-ai-copyright-notice-and-web-crawlers/}, September 8 2023{\natexlab{b}}.

\bibitem[Guild(2023{\natexlab{c}})]{authors2023survey}
The~Authors Guild.
\newblock Survey reveals 90 percent of writers believe authors should be compensated for the use of their books in training generative {AI}.
\newblock \url{https://authorsguild.org/news/ai-survey-90-percent-of-writers-believe-authors-should-be-compensated-for-ai-training-use/}, May 15 2023{\natexlab{c}}.

\bibitem[in~Fiction(2023)]{humanity2023authors}
Humanity in~Fiction.
\newblock Authors \& editors survey 2023.
\newblock \url{https://docsend.com/view/2v2sytxjey83nd4b}, 2023.

\bibitem[Ippolito et~al.(2022)Ippolito, Yuan, Coenen, and Burnam]{ippolito2022creative}
Daphne Ippolito, Ann Yuan, Andy Coenen, and Sehmon Burnam.
\newblock Creative writing with an {AI}-powered writing assistant: Perspectives from professional writers.
\newblock \emph{arXiv preprint arXiv:2211.05030}, 2022.

\bibitem[Jiang et~al.(2023)Jiang, Brown, Cheng, Khan, Gupta, Workman, Hanna, and Gebru]{jiang2023ai}
Harry~H. Jiang, Lauren Brown, Jessica Cheng, Mehtab Khan, Abhishek Gupta, Deja Workman, Jonathan Hanna, Alex an~Flowers, and Timnit Gebru.
\newblock {AI} art and its impact on artists.
\newblock In \emph{AAAI/ACM Conference on AI, Ethics, and Society}, Montreal, QC, August 2023.

\bibitem[Kaila et~al.(2024)Kaila, Holzapfel, and J{\"a}{\"a}skel{\"a}inen]{kaila2024gardening}
Anna-Kaisa Kaila, Andr{\'e} Holzapfel, and Petra J{\"a}{\"a}skel{\"a}inen.
\newblock Gardening frictions in creative {AI:} emerging art practices and their design implications.
\newblock In \emph{Proceedings of the 15th International Conference on Computational Creativity}, J{\"o}nk{\"o}ping, Sweden, June 2024.

\bibitem[Khan(2016)]{khan2016amazon}
Lina~M Khan.
\newblock Amazon's antitrust paradox.
\newblock \emph{Yale Law Journal}, 126:\penalty0 710, 2016.

\bibitem[Kirchenbauer et~al.(2023)Kirchenbauer, Geiping, Wen, Katz, Miers, and Goldstein]{kirchenbauer2023watermark}
John Kirchenbauer, Jonas Geiping, Yuxin Wen, Jonathan Katz, Ian Miers, and Tom Goldstein.
\newblock A watermark for large language models.
\newblock In \emph{International Conference on Machine Learning}, pages 17061--17084. PMLR, 2023.

\bibitem[Lemley and Casey(2020)]{lemley2020fair}
Mark~A Lemley and Bryan Casey.
\newblock Fair learning.
\newblock \emph{Texas Law Review}, 99:\penalty0 743, 2020.

\bibitem[Licensing and Society(2022)]{authors2022why}
Authors' Licensing and Collecting Society.
\newblock Why writers are at a loss for words.
\newblock \url{https://www.alcs.co.uk/news/why-writers-are-at-a-loss-for-words/}, December 6 2022.

\bibitem[Magazine(2024)]{uncanny2024please}
Uncanny Magazine.
\newblock {Please SHARE! Writers! We want to apologize...} [social media post].
\newblock \url{https://bsky.app/profile/uncannymagazine.bsky.social/post/3l6675ofnyn22}, October 10 2024.

\bibitem[Mesler et~al.(2024)Mesler, Howie, Chernishenko, Shen, and Vredenburg]{mesler2024association}
Rhiannon~M. Mesler, Katharine Howie, Jennifer Chernishenko, Mingnan~Nancy Shen, and Jessica Vredenburg.
\newblock The association between political identity centrality and cancelling proclivity.
\newblock \emph{Acta Psychologia}, 244, 2024.

\bibitem[Portsmouth(2024)]{portsmouth2024storywise}
Amy Portsmouth.
\newblock Storywise and open submissions {FAQ's}.
\newblock \url{https://angryrobotbooks.com/2024/04/open-submissions-2024/}, April 8 2024.

\bibitem[Redaelli(2024)]{redaelli2024intentionality}
Roberto Redaelli.
\newblock Intentionality gap and preter-intentionality in generative artificial intelligence.
\newblock \emph{AI \& SOCIETY}, pages 1--8, 2024.

\bibitem[Reisner(2023)]{reisner2023revealed}
Alex Reisner.
\newblock Revealed: The authors whose pirated books are powering generative {AI}.
\newblock \emph{The Atlantic}, Aug 19 2023.

\bibitem[Shan et~al.(2024)Shan, Ding, Passananti, Wu, Zheng, and Zhao]{shan2024nightshade}
Shawn Shan, Wenxin Ding, Josephine Passananti, Stanley Wu, Haitao Zheng, and Ben~Y Zhao.
\newblock Nightshade: Prompt-specific poisoning attacks on text-to-image generative models.
\newblock In \emph{2024 IEEE Symposium on Security and Privacy (SP)}, pages 212--212. IEEE Computer Society, 2024.

\bibitem[Stewart(2024)]{stewart2024tor}
Sophia Stewart.
\newblock To books criticized for use of {AI-}generated art in {'Gothikana'} cover design.
\newblock \emph{Publishers Weekly}, Feb 21 2024.

\bibitem[Tapper(2023)]{tapper2023authors}
James Tapper.
\newblock Authors shocked to find {AI} ripoffs of their books being sold on {Amazon}.
\newblock \emph{The Guardian}, Sep 30 2023.

\bibitem[Tobler(2024)]{tobler2024deadlands}
E.~Catherine Tobler.
\newblock {The Deadlands} also saw this during our last open period... [social media post].
\newblock \url{https://bsky.app/profile/ect.bsky.social/post/3l6dtdarc7n2x}, October 12 2024.

\bibitem[Tolbert(2019)]{tolbert2019sfwa}
Jeremiah Tolbert.
\newblock {SFWA} minimum pro rate now in effect.
\newblock \emph{Science Fiction \& Fantasy Writers' Association Blog}, September 3 2019.

\bibitem[Uusitalo et~al.(2024)Uusitalo, Salovaara, Jokela, and Salmimaa]{uusitalo2024clay}
Severi Uusitalo, Antti Salovaara, Tero Jokela, and Marja Salmimaa.
\newblock `{Clay} to play with': Generative {AI} tools in {UX} and industrial design practice.
\newblock In \emph{Proceedings of the 2024 ACM Designing Interactive Systems Conference}, pages 1566--1578, 2024.

\bibitem[Vimpari et~al.(2023)Vimpari, Kultima, H{\"a}m{\"a}l{\"a}inen, and Guckelsberger]{vimpari2023adapt}
Veera Vimpari, Annakaisa Kultima, Perttu H{\"a}m{\"a}l{\"a}inen, and Christian Guckelsberger.
\newblock {"An Adapt-or-Die Type of Situation"}: Perception, adoption, and use of text-to-image generation {AI} by game industry professionals.
\newblock In \emph{Proceedings of the ACM on Human-Computer Interaction}, volume~7, pages 131--164, 2023.

\bibitem[Wan et~al.(2023)Wan, Wallace, Shen, and Klein]{wan2023poisoning}
Alexander Wan, Eric Wallace, Sheng Shen, and Dan Klein.
\newblock Poisoning language models during instruction tuning.
\newblock In \emph{International Conference on Machine Learning}, pages 35413--35425. PMLR, 2023.

\bibitem[Wang et~al.(2024)Wang, Deng, Chiba-Okabe, Barak, and Su]{wang2024economic}
Jiachen~T Wang, Zhun Deng, Hiroaki Chiba-Okabe, Boaz Barak, and Weijie~J Su.
\newblock An economic solution to copyright challenges of generative ai.
\newblock \emph{arXiv preprint arXiv:2404.13964}, 2024.

\bibitem[Weatherbed(2023)]{weatherbed2023not}
Jess Weatherbed.
\newblock Not even {NYT} bestsellers are safe from {AI} cover art.
\newblock \emph{The Verge}, May 15 2023.

\bibitem[Writers(2023)]{international2023artificial}
International~Thriller Writers.
\newblock Artificial intelligence survey.
\newblock \url{https://thrillerwriters.org/artificial-intelligence-survey/}, March 2023.

\end{thebibliography}

\end{document}